\documentstyle[sprocl,epsf,12pt]{article}

\epsfverbosetrue
\newcommand{\cc}{c\bar{c}}
\newcommand{\bb}{b\bar{b}}

\newcommand{\al}{\alpha}
\newcommand{\GeV}{{\rm GeV}}

\newcommand{\ms}{\overline{{\rm MS}}}
\newcommand{\Tr}{{\rm Tr}}

\newcommand{\be}{\begin{equation}}
\newcommand{\ee}{\end{equation}}
\newcommand{\gtap}{{\raise.3ex\hbox{$>$\kern-.75em\lower1ex\hbox{$\sim$}}}}
\newcommand{\ltap}{{\raise.3ex\hbox{$<$\kern-.75em\lower1ex\hbox{$\sim$}}}}

\newcommand{\PRD}[3]{{\em Phys. Rev.} {\bf D{#1}} (19{#2}) {#3}}

\newcommand{\PRL}[3]{{\em Phys. Rev. Lett.} {\bf {#1}} (19{#2}) {#3}}
\newcommand{\PLB}[3]{{\em Phys. Lett.} {\bf B{#1}} (19{#2}) {#3}}
\newcommand{\NPB}[3]{{\em Nucl. Phys.} {\bf B{#1}} (19{#2}) {#3}}
\newcommand{\NPBproc}[3]{{\em Nucl. Phys.} {\bf B} (Proc. Suppl.)
           {\bf {#1}} (19{#2}) {#3}}
\newcommand{\ARNPS}[3]{{\em Annu. Rev. Nucl. Part. Sci.} 
           {\bf #1} (19{#2}) {#3}}

\begin{document}
\renewcommand{\thefootnote}{\fnsymbol{footnote}}
\begin{flushright}
ILL-TH-96-05  \\
hep-ph/9608220
\end{flushright}
\ \ \vskip3cm
\title{\bf $\alpha_s$ WITH LATTICE QCD\footnote{Presented at the 
XXXIst Rencontres de Moriond on Electroweak Interactions 
and Unified Theories, Les Arcs 1800, France, March 16-23, 1996}}

\author{Aida X. EL-KHADRA}

\address {Department of Physics,University of Illinois,\\ 1110 West Green
Street \\  Urbana, IL, 61801-3080}

\maketitle\vskip4in
\abstracts{
The status of determinations of the strong coupling constant based on lattice 
QCD is reviewed. }

\newpage
\baselineskip=18pt

\setcounter{footnote}{0}
\renewcommand{\thefootnote}{\alph{footnote}}

\section{Introduction and Motivation} \label{sec:IM}

\indent\indent
At present, the QCD coupling, $\al_s$, is determined from many 
different experiments, performed at energies ranging from a few 
to more than 100 GeV.\cite{als_rev}$^)$ In most cases perturbation theory is used to
extract $\al_s$ from the experimental information.
Experimental and theoretical progress over the last few years has 
made these determinations increasingly precise. 
However, all determinations, including those based on lattice QCD,
rely on phenomenologically-estimated
corrections and uncertainties from non-perturbative effects.
These effects will eventually (or already do) limit the accuracy 
of the coupling constant determination.
When lattice QCD is used the limiting uncertainty 
comes from the (total or partial) omission of sea quarks in numerical 
simulations.

The determination of the strong coupling, $\al_s$, proceeds in
three steps:

\begin{enumerate}

\item
The first step is always an experimental measurement.
In $\alpha_s$ determinations based on perturbative QCD this might be a 
cross section or (ratio of) decay rates. In determinations based on lattice
QCD this is usually a hadron mass or mass splitting, for example the mass
of the $\rho$ meson, or a better choice, spin-averaged splittings in the
charmonium and bottomonium system. In "lattice language" this step is often
referred to as "setting the scale" (see section~\ref{sec:scale}).

\item
The second step involves a choice of renormalization 
scheme. In perturbative QCD the standard choice is the $\ms$ scheme, which 
is only defined perturbatively. With lattice QCD a non-perturbative 
scheme may be choosen, and there are many candidates. In order to compare
with perturbative QCD, any such scheme should be accessible to perturbative
calculations (without excessive effort).

\item
Finally, the third step is an assessment of the 
experimental and theoretical errors associated with the strong coupling
determination. This is of course the most important (and sometimes also the
most controversial) step as it allows us to distinguish and weight different
determinations. As a theorist, I don't have much to say about experimental 
errors, other than that they should be small and controlled. The 
experimental errors on hadron masses are negligibly small in lattice 
determinations of $\alpha_s$ at this point.
The theoretical errors that are part of $\alpha_s$ determinations based on
perturbative QCD include higher order terms in the truncated perturbative 
series and the associated dependence on the renormalization scale, and 
hadronization or other generic non-perturbative effects. In lattice QCD
the theoretical errors include (but are not limited to) discretization 
errors (due to the finite lattice spacing, $a\neq 0$), finite volume 
effects, and errors associated with the partial or total omission of sea
quarks.

\end{enumerate}

The consideration of systematic uncertainties should guide us towards quantities
where these uncertainties are controlled, for a reliable determination of $\alpha_s$. 
As has been argued by Lepage,\cite{lepage}$^)$ quarkonia are among
the easiest systems to study with lattice QCD, since systematic
errors can be analyzed easily  with potential models if not by
brute force.

Finite-volume errors are much easier to control for
quarkonia than for light hadrons, since quarkonia are smaller.
Lattice-spacing errors, on the other hand, can be larger
for quarkonia and need to be considered.
An alternative to reducing the lattice spacing in order to control
this systematic error is improving the action (and operators).
For quarkonia, the size of lattice-spacing errors in a numerical 
simulation can be {\em anticipated} by calculating expectation 
values of the corresponding operators using potential-model wave 
functions. They are therefore ideal systems to test and establish
improvement techniques.

A lot of the work of phenomenological relevance is done in
what is generally referred to as the ``quenched'' 
(and sometimes as the ``valence'') approximation.
In this approximation gluons are not allowed to split into 
quark - anti-quark pairs (sea quarks). In the case of
quarkonia, potential model phenomenology can be used to
estimate this systematic error.

I shall have neither the time nor the space to give an introduction to
lattice QCD. Instead, I refer the reader to a number 
of pedagogical introductions and reviews in the literature.\cite{intro}$^)$

\section{Determination of the Lattice Spacing and the Quarkonium Spectrum}
   \label{sec:scale}

\begin{figure}[htb]
\begin{center}
\epsfxsize= 0.65\textwidth
\leavevmode
\epsfbox{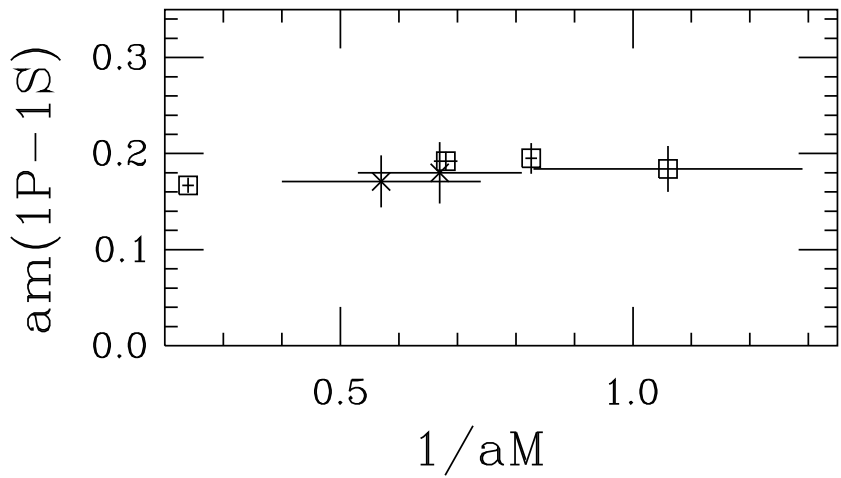}
\end{center}
\caption{The 1P-1S splitting as a function of the 1S mass
   (statistical errors only) from Ref. [5];
   $\Box$: ${\cal O}(a^2)$ errors; $\times$: ${\cal O}(a)$ errors.}
 \label{fig:1p1s}
\end{figure}

\indent\indent
The experimental input to the strong coupling determination is
a mass or mass splitting, from which by comparison with the
corresponding lattice quantity the lattice spacing, $a$, is determined in 
physical units.
For this purpose, one should identify quantities that are insensitive 
to lattice errors. In quarkonia, spin-averaged splittings are good
candidates. The experimentally observed 1P-1S and 2S-1S splittings
depend only mildly on the quark mass (for masses between $m_b$ and $m_c$).
Figure~\ref{fig:1p1s} 
shows the observed mass dependence of the 1P-1S splitting
in a lattice QCD calculation. The comparison between results from
different lattice actions illustrates that
higher-order lattice-spacing errors for these splittings
are small.\cite{nrqcd_als,us}$^)$

Two different formulations for fermions have been used in lattice 
calculations of the quarkonia spectra.
In the non-relativistic limit the QCD action can be written as
an expansion in powers of $v^2$ (or $1/m$), where $v$ is the 
velocity of the heavy quark inside the boundstate;\cite{eff}$^)$
I shall henceforth refer to this approach as NRQCD. Lepage and
collaborators \cite{nrqcd_thy}$^)$ have adapted this formalism to
the lattice regulator. Several groups have performed numerical
calculations of quarkonia in this approach.
In \mbox{Refs.~[8,9]} the 
NRQCD action is used to calculate
the $\bb$ and $\cc$ spectra, including 
terms up to ${\cal O} (mv^4)$ and ${\cal O}(a^2)$.
In addition to calculations in the quenched approximation,
this group is also using gauge configurations that include
2 flavors of sea quarks with mass $m_q \sim \frac{1}{2} m_s$ 
to calculate the $\bb$ spectrum.\cite{nrqcd_als,shige}$^)$
The leading order NRQCD action is used in Ref.~[11] 
for a calculation of the $\bb$ spectrum in the quenched approximation.

The Fermilab group \cite{us_thy}$^)$ developed a generalization of previous 
approaches, which encompasses the non-relativistic
limit for heavy quarks as well as Wilson's relativistic action
for light quarks. Lattice-spacing artifacts are analyzed for quarks with
arbitrary mass. Ref.~[5] uses this approach to calculate
the $\bb$ and $\cc$ spectra in the quenched approximation. We
considered the effect of reducing lattice-spacing errors from
${\cal O}(a)$ to ${\cal O}(a^2)$. 
The SCRI collaboration \cite{sloan}$^)$ is also using this approach 
for a calculation of the $\bb$ spectrum using the same gauge configurations 
as the NRQCD collaboration with $n_f = 2$ and an improved fermion action
(with ${\cal O}(a^2)$ errors).

\begin{figure}[htb]
\begin{center}
\epsfxsize= 0.7\textwidth
\leavevmode
\epsfbox[4 164 595 614]{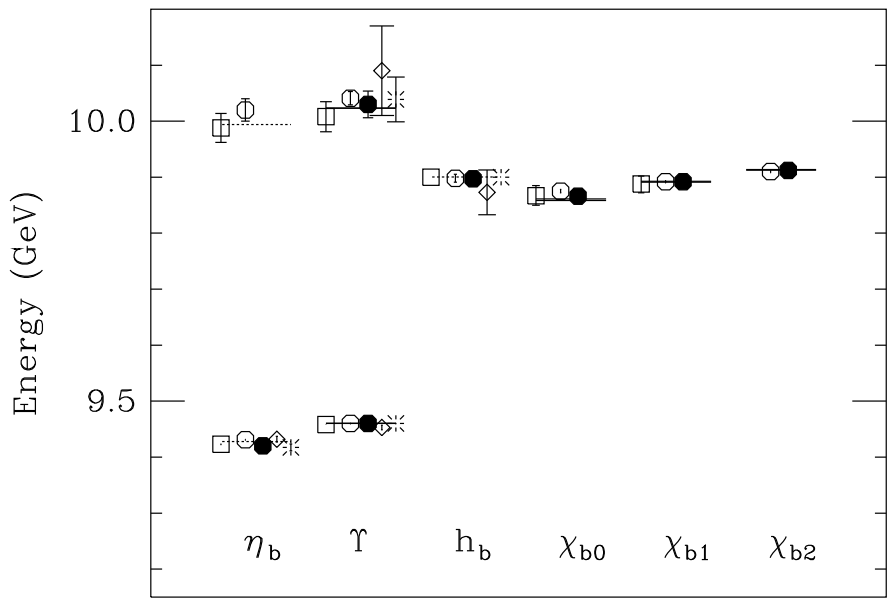}
\end{center}
\caption{A comparison of lattice QCD results for the $\bb$ spectrum
   (statistical errors only).
   -: Experiment; $\Box$: FNAL [5]; $\circ$: NRQCD ($n_f=0$) [8];
   $\bullet$: NRQCD ($n_f=2$) [4]; $\Diamond$: UK(NR)QCD [11];
   $*$: SCRI [13].}
   \label{fig:bb}
\end{figure}

\begin{figure}[htb]
\begin{center}
\epsfxsize= 0.65\textwidth
\leavevmode
\epsfbox[4 164 595 614]{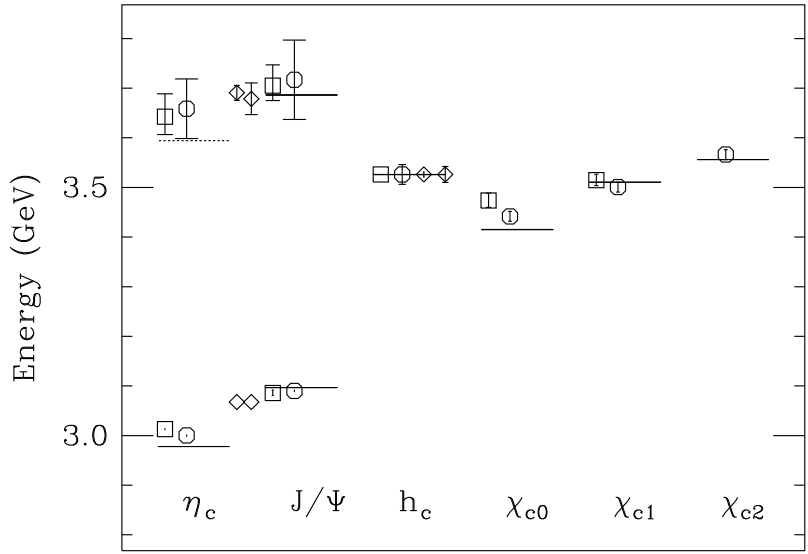}
\end{center}
\caption[xxx]{A comparison of lattice QCD results for the $\cc$ spectrum
   (statistical errors only).
   -: Experiment; $\Box$: FNAL [5]; $\circ$: NRQCD ($n_f=0$) [9];
   $\Diamond$: ADHLM [14].}
 \label{fig:cc}
\end{figure}

All but one group use gauge configurations generated with the Wilson 
action, leaving ${\cal O}(a^2)$ lattice-spacing errors in the
results. The lattice spacings, in this case, are in the range
$a \simeq 0.05 - 0.2$ fm.
Ref.~[14] uses an improved gauge action together with
a non-relativistic quark action improved to the same order
(but without spin-dependent terms) on coarse ($a \simeq 0.4 - 0.24$ fm)
lattices.
The results for the $\bb$ and $\cc$ spectra from all groups are 
summarized in Figures~\ref{fig:bb}~and~\ref{fig:cc}.

The agreement between the experimentally-observed spectrum and 
lattice QCD calculations is impressive. As indicated in the
preceding paragraphs, the lattice artifacts are different for all groups. 
Figures~\ref{fig:bb}~and~\ref{fig:cc} therefore emphasize the level of 
control over systematic errors.

Results with 2 flavors of degenerate sea quarks
have now become available from a number of 
groups,\cite{nrqcd_als,kek_als,cdhw,shige}$^)$ with lattice-spacing 
and finite-volume errors similar to the quenched calculations, 
significantly reducing this systematic error.
However, several systematic effects associated with the 
inclusion of sea quarks still need to be studied further.
They include the dependence of the quarkonium
spectrum on the number of flavors of sea quarks, and 
the sea-quark action (staggered vs. Wilson). The inclusion
of sea quarks with realistic light-quark masses is very
difficult. However, quarkonia are expected to depend only 
very mildly on the masses of the light quarks.

\section{Definition of a Renormalized Coupling}

\indent\indent
Within the framework of lattice QCD the conversion from the bare
to a renormalized coupling can, in principle, be made
non-perturbatively. In the definition of a renormalized coupling, 
systematic uncertainties should be controllable, and at short 
distances, its (perturbative) relation
to other conventional definitions calculable.
For example, a renormalized coupling can
be defined from the non-perturbatively computed heavy-quark potential 
($\al_V$).\cite{schill}$^)$
An elegant approach has been developed in Ref.~[18], where
a renormalized coupling is defined non-perturbatively through 
the Schr\"{o}dinger functional. The authors compute the evolution
of the coupling non-perturbatively using a finite size 
scaling technique, which allows them to vary the momentum
scales by an order of magnitude.
The same technique has also been applied to the renormalized coupling 
defined from twisted Polyakov loops.\cite{alpha}$^)$ The numerical 
calculations include only gluons at the moment. However, the inclusion of
fermions is possible. Once such simulations become available they should
yield very accurate information on $\al_s$ and its evolution.
The strong coupling can also be computed from the three-gluon
vertex, suitably defined on the lattice.\cite{parr}$^)$

An alternative is to define a renormalized coupling through short
distance lattice quantities, like small Wilson loops or Creutz 
ratios which can be calculated perturbatively and by numerical simulation. For 
example, the coupling defined from the plaquette,  
$\al_P = - 3 \ln{ \langle \Tr \, U_P \rangle } / 4 \pi$,
can be expressed in terms of $\al_V$ (or $\al_{\ms}$) by:\cite{lm}$^)$
\begin{equation} \label{eq:lm19}
 \al_P = \al_V (q) [ 1 - (1.19  + 0.07 \, n_f) \al_V (q) + {\cal O} (\al_V^2) ]
\end{equation}
at $q=3.41/a$, close to the ultraviolet cut-off. 
$\al_V$ is related to the more commonly used $\ms$ coupling by 
\be    \label{eq:vms}
\al_{\ms} (Q) = \al_V (e^{5/6} Q) (1 + \frac{2}{\pi} \al_V + \ldots)
\;\;\;.
\ee
The size of higher-order corrections associated with the above 
defined coupling constants can be tested by comparing 
perturbative predictions for short-distance lattice quantities
with non-perturbative results.\cite{lm}$^)$
The comparison of the non-perturbatively calculated coupling of Ref.~[18]
with the perturbative predictions for this coupling using Eq.~(\ref{eq:lm19})
is an additional consistency test.

The relation of the plaquette coupling in Eq.~(\ref{eq:lm19}) 
to the $\ms$ coupling has recently been calculated to 2-loops 
\cite{alles,lw-2l}$^)$
in the quenched  approximation (no sea quarks, $n_f = 0$).
The extension to $n_f \neq 0$ will significantly reduce
the uncertainty due to the use of perturbation theory.

\section{Sea Quark Effects}   \label{sec:sea}

\indent\indent
Calculations that properly include all sea-quark effects 
do not yet exist.
If we want to make contact with the ``real world'', these effects 
have to be estimated phenomenologically or extrapolated away. 

The phenomenological correction necessary to account for
the sea-quark effects omitted in calculations of quarkonia
that use the quenched approximation gives rise to the dominant
systematic error in this calculation.\cite{prl,nrqcd_l93}$^)$
By demanding that, say, the spin-averaged 1P-1S splitting calculated on
the lattice reproduce the experimentally observed one (which
sets the lattice spacing, $a^{-1}$, in physical units), the effective
coupling of the quenched potential is in effect matched to the 
coupling of the effective 3 flavor potential at the typical
momentum scale of the quarkonium states in question. The difference
in the evolution of the zero flavor and 3,4 flavor couplings 
from the effective low-energy scale to the ultraviolet cut-off, 
where $\al_s$ is determined, is the perturbative estimate
of the correction.

For comparison with other determinations of $\al_s$, the $\ms$ 
coupling can be evolved to the $Z$ mass scale. An average 
\cite{als_rev}$^)$ of 
Refs.~[24,25] yields for $\al_s$ from calculations
in the quenched approximation:
\be          \label{eq:nf0}
   \al^{(5)}_{\ms} (m_Z) = 0.110 \pm 0.006 \;\;\;.
\ee

The phenomenological correction described in the previous paragraph
has been tested from first principles in Ref.~[15].
The 2-loop evolution of $n_f = 0$ and $n_f = 2$ $\ms$ couplings 
-- extracted from calculations of the $\cc$ spectrum using the 
Wilson action in the quenched approximation and with 2 flavors of 
sea quarks respectively -- to the low-energy scale gives consistent
results. After correcting the 2 flavor result to $n_f = 3$ in the
same manner as before and evolving $\al_{\ms}$ to the $Z$ mass,
they find \cite{kek_als}$^)$
\be
 \al^{(5)}_{\ms} (m_Z) = 0.111 \pm 0.005 
\ee
in good agreement with the previous result in Eq.~(\ref{eq:nf0}).
The total error is now dominated by the rather large 
statistical errors and the perturbative uncertainty.

The most accurate result to date comes from the NRQCD 
collaboration.\cite{nrqcd_als,shige}$^)$
They use results for $\al_s$ from the $\bb$
spectrum with 0 and 2 flavors of sea quarks to extrapolate
the inverse coupling to the physical 3 flavor case directly at
the ultraviolet momentum, $q = 3.41/a$. They obtain a result 
consistent with the old procedure.
Recently, they have begun to study the dependence of $\al_s$ on
the masses of the sea quarks. Their preliminary result is:
\be   \label{eq:v3}
    \al_V^{(3)} (8.2 \, \GeV) = 0.195 \pm 0.003 \pm 0.001 \pm 0.004 \;\;\;.
\ee
The first error is statistics, the second error an estimate of 
residual cut-off effects and the third (dominant) error is due to the 
quark mass dependence.
The conversion to $\ms$ (including the 2-loop term of Refs.~[22,23]) and 
evolution to the $Z$ mass then gives:
\be        \label{eq:ms5}
\al^{(5)}_{\ms} (m_Z) = 0.118 \pm 0.003 \;\;\;,
\ee
where the error now also includes the perturbative uncertainty from 
eq.~(\ref{eq:vms}). A similar analysis is performed in Ref.~[16]
on the same gauge configurations but using the Wilson action for
a calculation of the $\cc$ spectrum. The result for the coupling
is consistent with Refs.~[4,15].

The preliminary calculation of the SCRI collaboration\cite{sloan}$^)$ 
($n_f=2$) can be combined with the result of Ref.~[5]. Using the same analysis 
as in Ref.~[4] gives \cite{shige}$^)$
\be \label{eq:ms5n}
\al^{(5)}_{\ms} (m_Z) = 0.116 \pm 0.003 \;\;,
\ee
nicely consistent with eqn.~(\ref{eq:ms5}).
Clearly, more work is needed to confirm the results of eqns.~(\ref{eq:ms5})
and (\ref{eq:ms5n}), especially in calculations with heavy quark actions
based on Ref.~[12].
In particular, the systematic errors associated with the 
inclusion of sea quarks into the simulation have to be checked,
as outlined in section~\ref{sec:scale}.

\section{Conclusions} \label{sec:con}

\indent\indent
Phenomenological corrections are a necessary evil that enter most
coupling constant determinations.
In contrast, lattice QCD calculations with complete control over
systematic errors will yield truly first-principles determinations 
of $\al_s$ from the experimentally observed hadron spectrum.

At present, determinations of $\al_s$ from the experimentally measured
quarkonia spectra using lattice
QCD are comparable in reliability and accuracy to other determinations
based on perturbative QCD from high energy experiments. They
are therefore part of the 1995 world average for $\al_s$.\cite{als_rev}$^)$
The phenomenological corrections for the most important sources
of systematic errors in lattice QCD calculations of quarkonia have already
been replaced by first principles calculations. This has led to a 
significant increase in the accuracy of $\al_s$ determinations from 
quarkonia.

Still lacking for a first-principles result is the
proper inclusion of sea quarks. A difficult problem in this context 
is the inclusion of sea quarks with physical light quark masses. At present, 
this can only be achieved by extrapolation (from $m_q \simeq 0.3 - 0.5 m_s$
to $m_{u,d}$).
If the light quark mass dependence of the quarkonia spectra
is mild, as indicated by the NRQCD collaboration, the associated systematic 
error can be controlled.
First-principles calculations of quarkonia could then be
performed with currently available computational resources. 

\section*{Acknowledgements}

I thank the organizers for an enjoyable conference and J. Shigemitsu, 
J. Simone, and J. Sloan for assistance in preparing this talk.

\setlength{\baselineskip}{0.8\baselineskip}

\end{document}